\documentclass[%
 reprint,
superscriptaddress,
 amsmath,amssymb,
 aps,
]{revtex4-2}

\usepackage{graphicx}
\usepackage{dcolumn}
\usepackage{bm}
\usepackage{color}
\usepackage{aas_macros}
\usepackage{mathcomp}
\usepackage{comment}
\usepackage[scr=boondox]{mathalpha}
\usepackage[normalem]{ulem}

\let\originalcite\cite
\let\originalcitet\citet
\let\originalcitep\citep

\NewDocumentCommand{\trackedcite}{o o m}{%
  \mbox{%
    \IfNoValueTF{#1}
      {\originalcite{#3}}
      {\IfNoValueTF{#2}
        {\originalcite[#1]{#3}}
        {\originalcite[#1][#2]{#3}}}%
  }%
}

\NewDocumentCommand{\trackedcitet}{o o m}{%
  \mbox{%
    \IfNoValueTF{#1}
      {\originalcitet{#3}}
      {\IfNoValueTF{#2}
        {\originalcitet[#1]{#3}}
        {\originalcitet[#1][#2]{#3}}}%
  }%
}

\NewDocumentCommand{\trackedcitep}{o o m}{%
  \mbox{%
    \IfNoValueTF{#1}
      {\originalcitep{#3}}
      {\IfNoValueTF{#2}
        {\originalcitep[#1]{#3}}
        {\originalcitep[#1][#2]{#3}}}%
  }%
}

\NewDocumentCommand{\withprotectedcitations}{+m}{%
  \begingroup
  \let\cite\trackedcite
  \let\citet\trackedcitet
  \let\citep\trackedcitep
  #1%
  \endgroup
}

\begin{document}

\preprint{APS/123-QED}

\title{Inferring population III star properties from the 21-cm global signal}

\author{Sho Ukai}
\email{ukai.sho.x7@s.mail.nagoya-u.ac.jp}
\affiliation{%
 Graduate School of Science, Nagoya University, Furocho, Chikusa-ku, Nagoya, Aichi, 464-8602, Japan
}%

\author{Hayato Shimabukuro}
\affiliation{%
 Graduate School of Science, Nagoya University, Furocho, Chikusa-ku, Nagoya, Aichi, 464-8602, Japan
}%
\affiliation{%
 South-Western Institute for Astronomy Research (SWIFAR), Yunnan University, Kunming, Yunnan 650500, People's Republic of China
}%
\affiliation{%
 Key Laboratory of Survey Science of Yunnan Province, Yunnan University, Kunming, Yunnan 650500, People's Republic of China
}%

\author{Kenji Hasegawa}
\affiliation{%
 Department of Mechanical Engineering, National Institute of Technology, Suzuka College, Shiroko-cho, Suzuka, Mie, 510-0294, Japan
}%

\author{Kiyotomo Ichiki}
\affiliation{%
 Graduate School of Science, Nagoya University, Furocho, Chikusa-ku, Nagoya, Aichi, 464-8602, Japan
}%
\affiliation{%
 Kobayashi-Maskawa Institute for the Origin of Particles and the Universe, Nagoya University, Furocho, Chikusa-ku, Nagoya, Aichi, 464-8602, Japan
}%
\affiliation{%
 Institute for Advanced Research, Nagoya University, Furocho, Chikusa-ku, Nagoya, Aichi 464-8602, Japan
}%

\date{\today}

\begin{abstract}
Investigating the properties of the first stars in the universe, known as Population III (Pop~III) stars, is essential, yet it remains an open question. One way to explore these stars is by examining their effects on the surrounding gas during the cosmic dawn. In this study, we investigate whether the 21-cm global signal can constrain the typical mass and star formation efficiency of first-generation stars. We perform semi-numerical simulations that include the escape fraction of ionizing photons, which depends on stellar and halo masses, as well as the heating structure surrounding a halo that hosts the first star, determined by radiation hydrodynamics (RHD) simulations.
Using a Fisher analysis that includes thermal noise for an integration time of $1000\,\mathrm{h}$, we find that the global signal provides information for constraining these properties if the foreground spectrum can be perfectly removed.
However, when the smooth foreground spectrum is modeled using a log-polynomial and inferred from the same data, its spectral variation becomes strongly degenerate with the changes produced by the Pop~III parameters, substantially weakening the constraints.
These results demonstrate that accurate foreground modeling and removal are essential for extracting information about Pop~III properties from the global 21-cm signal.
\end{abstract}

\maketitle


\section{Introduction}
Observations of the Cosmic Microwave Background (CMB) reveal a nearly homogeneous Universe about 380,000 years after the Big Bang, consisting primarily of hydrogen and helium. Subsequently, gravitational collapse initiated the formation of the first generation of stars (Population III; Pop~III) in the universe from pristine gas clouds devoid of metals. Pop~III stars fundamentally altered the state of the universe by generating the first heavy elements and emitting intense ultraviolet (UV) radiation, thus ending the cosmic "dark ages" and initiating the cosmic dawn. The subsequent formation of metal-enriched Population II (Pop~II) stars produced abundant UV radiation and played an important role in driving the progress of the epoch of reionization (EoR) \cite{2011ARA&A..49..373B}.

Theoretical predictions suggest that Pop~III stars formed under different physical conditions compared to later generations, primarily due to less efficient gas cooling processes, resulting in more massive stars. Simulation studies indicate that Pop~III stars could have typical masses ranging from tens to hundreds of solar masses, significantly exceeding those of later-generation stars \cite{10.1093/mnras/stv044,Hosokawa_2016}. Their immense luminosities and brief lifetimes (a few million years) culminated typically in supernovae, enriching the interstellar medium and marking a transition to metal-enriched star formation (Population II) \cite{RevModPhys.85.809,2018MNRAS.475.4396J}.

Despite their importance, direct observational evidence for Pop~III stars remains difficult due to their occurrence at high redshifts in faint, early proto-galaxies, beyond the reach of current observational capabilities \cite{2011ARA&A..49..373B}.
Consequently, their properties—including initial mass function (IMF), star formation efficiency, and overall cosmic impact—remain uncertain. 
Indirect evidence, such as chemical abundances in ultra-metal-poor stars and early galaxies, currently constrains these characteristics \cite{RevModPhys.85.809,2018MNRAS.475.4396J}.
To overcome observational limitations, the 21-cm line of neutral hydrogen has been considered a promising probe of early star formation. The global 21-cm signal, i.e., the global average of the redshifted 21-cm signal, traces the thermal and ionization history of the intergalactic medium (IGM) \cite{2006PhR...433..181F,2012RPPh...75h6901P,2016arXiv160301961H,2023PASJ...75S...1S}. 

The formation of the first stars at redshifts $z \gtrsim 20 - 30$ produced Ly$\alpha$ radiation that coupled the hydrogen spin temperature to the kinetic gas temperature, which is colder than the CMB temperature, via the Wouthuysen-Field effect, thereby manifesting as absorption features in the global 21-cm signal. Subsequent X-ray emission from early stellar populations heated the IGM, eventually shifting the signal from absorption to emission at $z \sim 15$ \cite{2006MNRAS.371..867F}. Ultimately, progressive reionization eliminated neutral hydrogen, extinguishing the 21-cm signal by $z \sim 6$, as shown from the analyses of Lyman-$\alpha$ forest \cite{2015MNRAS.447..499M,2023ApJ...942...59J}. Lyman alpha emitter statistics also show that the IGM is already highly ionized by $z \approx 6$, with the major phase of reionization completed by $z \gtrsim 7$ \cite{2010ApJ...723..869O,2014ApJ...797...16K,2018PASJ...70S..16K}.

Observations at higher redshift $z>10$ remain inconclusive. 
In 2018, the EDGES experiment reported a controversial global 21-cm absorption feature centered at $\sim 78$ MHz ($z \approx 17$), considerably deeper than standard astrophysical predictions \cite{2018Natur.555...67B}. 
Subsequent analyses questioned the cosmological interpretation due to potential calibration and foreground modeling issues \cite{2018Natur.564E..32H}. 
Independent measurements, including those from SARAS 3, have not confirmed the EDGES signal, underscoring the experimental challenges involved, primarily foreground contamination \cite{2022NatAs...6..607S}.
A challenge for global-signal experiments is that the expected 21-cm signal and foreground emission are both spectrally smooth, making their separation based on spectral information difficult \cite{PhysRevD.87.043002}.
Several ground-based global-signal experiments are ongoing, including EDGES-3, the Radio Experiment for the Analysis of Cosmic Hydrogen (REACH), the Mapper of the IGM Spin Temperature (MIST), the Electromagnetically Isolated Global Signal Estimation Platform (EIGSEP), and Global Imprints from Nascent Atoms to Now (GINAN) \cite{2025PASP..137l5002C, 2022NatAs...6..984D, 2024MNRAS.530.4125M, 2026arXiv260202661B, 2026NatAs.tmp..122M}.
Meanwhile, interferometric arrays such as Low Frequency Array (LOFAR), Murchison Widefield Array (MWA), and Hydrogen Epoch of Reionization Array (HERA) aim to measure spatial fluctuations in the 21-cm signal, complementing global signal studies by providing additional constraints on reionization and heating timelines \cite{2013A&A...556A...2V, 2013PASA...30....7T, 2017PASP..129d5001D}.

The predicted 21-cm global signal \cite{2018MNRAS.478.5591M,2022MNRAS.516..841G,2025NatAs...9.1268G} and power spectrum \citep{2025MNRAS.540..483V} are highly sensitive to key Pop~III properties — including their typical stellar mass, formation efficiency, initial mass function, and overall cosmic impact — with variations in these quantities shifting the timing and modifying the amplitude of the associated absorption and emission features. The degeneracy among these astrophysical properties and foreground residuals complicates efforts to isolate distinct Pop~III signatures within the global 21-cm data. Therefore, it remains unclear to what extent future observations can uniquely constrain the characteristics of Pop~III stars.

This study aims to address these uncertainties by quantifying the feasibility of constraining the properties of Pop~III stars through detailed theoretical modeling and Fisher matrix analysis. 
Specifically, we assess how forthcoming global 21-cm observations, accounting for foreground contamination and thermal noise, can constrain Pop~III stellar parameters, namely the star formation efficiency ($f_\ast$) and the individual stellar mass $M_\mathrm{s}$.
We find below that $f_\ast$ and $M_\mathrm{s}$ can be constrained under
the idealized assumption of perfect foreground removal, whereas
marginalizing over the smooth foreground parameters substantially
weakens the constraints because of foreground--signal degeneracies.

The paper is organized as follows. In Section \ref{sec:impact of first star properties on 21-cm signal}, we describe the methodology for simulating the 21-cm signal depending on Pop~III properties and present the simulation results. In Section \ref{sec:fisher forecast}, we estimate the constraints on Pop~III properties from observations of the 21-cm global signal using a Fisher matrix analysis. In Section \ref{sec:discussion}, we discuss the implications and limitations of our results. In Section \ref{sec:summary}, we summarize our conclusions.

Throughout this paper, we work with a flat $\Lambda$CDM cosmology consistent with the Planck result: ($\Omega_m$, $\Omega_b$, $h$, $n_\mathrm{s}$, $\sigma_8$, $Y_\mathrm{He}$) = (0.315, 0.0493, 0.674, 0.965, 0.811, 0.245) \cite{2020A&A...641A...6P}, where $\Omega_{\rm m}$ and $\Omega_{\rm b}$ are the matter and baryon density parameters, respectively, $h$ is the normalized Hubble parameter, $n_s$ is the spectral index of the power spectrum of the primordial density fluctuation, $\sigma_8$ is the matter fluctuation amplitude at $8 h^{-1}$ Mpc, and $Y_{\rm He}$ is Helium mass fraction.

\section{impact of first star properties on 21-cm signal}
\label{sec:impact of first star properties on 21-cm signal}
In this section, after briefly reviewing how Pop~III properties affect the 21-cm signal, we describe the methodology for simulating the 21-cm signal as a function of Pop~III properties and present the simulation results.

The cosmological 21-cm signal, the offset of 21-cm brightness temperature from CMB temperature, is written as \cite{2012RPPh...75h6901P, 2010PhRvD..82b3006P}
\begin{equation}
\begin{aligned}
    \delta T_{b}(z) \approx & 27 x_\mathrm{HI} \left( 1+\delta_b \right) \left( \frac{\Omega_b h^2}{0.023} \right) \left( \frac{0.15}{\Omega_m h^2} \frac{1+z}{10} \right)^{1/2} \\ & \times\left( \frac{T_S - T_\mathrm{CMB}(z)}{T_S} \right) \left[\frac{(1+z) H(z)}{\partial_r v_r}\right] \mathrm{mK}~,
    \label{eq:Tb}
\end{aligned}
\end{equation}
where $x_\mathrm{HI}$, $\delta_b$, $T_\mathrm{S}$, $T_\mathrm{CMB}$, and $\partial_rv_r$ denote the fraction of neutral hydrogen in the intergalactic medium (IGM), the overdensity of baryon, the spin temperature of hydrogen, the CMB temperature at redshift $z$, and the velocity gradient of the line of sight, respectively. 
Although $\delta T_b$ has spatial variations, its sky-averaged value, known as the global signal, traces the overall evolution of the ionization and thermal state of the intergalactic medium (IGM). In this paper, we focus on this global signal. The spin temperature, $T_S$, is determined by
\begin{equation}
    T_S^{-1} = \frac{T_\mathrm{CMB}^{-1} + x_c T_K^{-1} + x_\alpha T_\alpha^{-1}}{1 + x_c + x_\alpha},
\end{equation}
where $T_K$, $T_\alpha$, $x_c$, and $x_\alpha$ are the gas kinetic temperature of hydrogen and the color temperature of Ly$\alpha$ photons, the collisional coupling constant, and Ly$\alpha$ coupling constant, respectively \cite{2012RPPh...75h6901P}.

At the end of the dark ages, the spin temperature \(T_S\) is coupled to the CMB temperature \(T_{\mathrm{CMB}}\) because the low gas density makes collisional coupling inefficient (\(x_c \ll 1\)), and the absence of stellar radiation means that Ly\(\alpha\) coupling is negligible. In this regime, 21\,cm transitions are mainly driven by the radiative coupling with the CMB, forcing \(T_S \simeq T_{\mathrm{CMB}}\) and producing no observable 21\,cm signal. 

After the birth of the first stars, the background of redshifted Ly\(\alpha\) photons activates the Wouthuysen--Field effect~\cite{1952AJ.....57R..31W,1959ApJ...129..536F}, which couples \(T_S\) to the gas kinetic temperature \(T_K\). In the standard cosmological scenario, \(T_K < T_{\mathrm{CMB}}\) at this stage, leading to an absorption feature in the global 21\,cm signal. As X-rays from the first sources subsequently heat the intergalactic medium, \(T_K\) (and thus \(T_S\)) rises and approaches \(T_{\mathrm{CMB}}\), causing the absorption trough to shallow and eventually vanish when \(T_S = T_{\mathrm{CMB}}\). 

Population III stars affect the 21-cm signal through several radiative processes. Ly$\alpha$ photons produced by Pop~III stars drive the Wouthuysen--Field coupling, while ionizing photons increase the ionized fraction and heat the IGM through photoionization. In addition, LW radiation suppresses subsequent Pop~III star formation by increasing the minimum halo mass required for gas cooling (for a review, see \cite{2023ARA&A..61...65K} and the references therein). Consequently, the frequency, depth, and shape of the global 21-cm absorption feature depend on the Pop~III star-formation efficiency and stellar mass.

To calculate the global 21-cm signal, we use the semi-numerical simulation code 21cmFAST \cite{2011MNRAS.411..955M} with the Pop~III module developed by \citet{2021MNRAS.502..463T}. 
The code generates the evolved density and velocity fields and calculates the ionization and thermal histories of the IGM, from which the spin temperature and 21-cm brightness temperature are obtained. 
The simulations are performed in a cubic volume with a side length of $256\,\mathrm{cMpc}$. 
A Gaussian random density field is generated on a $768^3$ grid from the linear matter power spectrum.
The density and velocity fields are then evolved from $z=300$ to each output redshift using the Zel'dovich approximation with second-order Lagrangian perturbation theory (2LPT) corrections.
The evolved density and velocity fields are constructed on a $256^3$ grid, corresponding to a spatial resolution of $1\,\mathrm{cMpc}$.
The ionization and thermal histories are also calculated on this grid.

Following \citet{2021MNRAS.502..463T}, we incorporate a Pop~III module that includes the stellar- and halo-mass-dependent escape fraction of ionizing photons, the cumulative production and recombination of ionizing photons, sub-grid UV photoheating around Pop~III-hosting halos, and LW feedback on the minimum halo mass for star formation. In the following subsections, we describe these components following the prescriptions of \citet{2021MNRAS.502..463T}.

\subsection{Local ionization around halos that host Pop~III stars}

The ionization field is constructed using the excursion-set filtering procedure employed in 21cmFAST \cite{2011MNRAS.411..955M}. In this procedure, the ionization state of each cell is evaluated using quantities spatially averaged over a scale $R$, which is varied from a maximum filtering scale, which will be defined below, down to the cell size.

For the ionizing-photon budget, we adopt the modified prescription developed by \citet{2021MNRAS.502..463T}. The standard 21cmFAST prescription determines the ionization state using the collapsed fraction evaluated at each redshift together with an ionizing efficiency. By contrast, the prescription adopted here explicitly follows the cumulative numbers of ionizing photons and recombinations over the previous evolution.

A cell is identified as fully ionized if the cumulative number of ionizing photons is sufficient to ionize the hydrogen and compensate for recombinations. The corresponding criterion is
\begin{equation}
    \label{eq:criterion}
    \bar{N}^R_\mathrm{ion}(\mathbf{x},z)
    >
    1+\bar{N}^R_\mathrm{rec}(\mathbf{x},z)~,
\end{equation}
where $\bar{N}^R_\mathrm{ion}(\mathbf{x},z)$ and $\bar{N}^R_\mathrm{rec}(\mathbf{x},z)$ are the cumulative numbers of ionizing photons and recombinations per baryon, respectively, spatially averaged over the scale $R$.

For each cell, the cumulative number of ionizing photons per baryon is calculated as
\begin{equation}
    N_\mathrm{ion}(\mathbf{x},z)
    =
    \int^{z_\mathrm{init}}_z
    \mathrm{d}z'\,
    \zeta_\mathrm{ion}(z')
    \frac{\mathrm{d}f_\mathrm{coll}(\mathbf{x},z')}
         {\mathrm{d}z'},
    \label{eq:Nion}
\end{equation}
where
\begin{equation}
    \zeta_\mathrm{ion}(z)
    =
    N_\mathrm{UV}f_\mathrm{esc}(z)f_\ast .
\end{equation}
Here, $N_\mathrm{UV}$ is the number of ionizing photons produced per stellar baryon. We adopt $N_\mathrm{UV}=70000$ following \citet{2021MNRAS.502..463T}. 
$f_\mathrm{esc}(z)$ is the escape fraction of ionizing photons, $f_\ast$ is the Pop~III star-formation efficiency, and $f_\mathrm{coll}(\mathbf{x},z)$ is the fraction of matter collapsed into halos capable of hosting Pop~III stars. 
The escape fraction $f_\mathrm{esc}(z)$ is described in Subsection~\ref{subsec:escapefraction}, while $f_\ast$ and $f_\mathrm{coll}$ are defined in Subsection~\ref{subsec:LW_feedback}.

The cumulative number of recombinations per baryon in each cell is calculated as
\begin{equation}
    N_\mathrm{rec}(\mathbf{x},z)
    =
    \int^{z_\mathrm{init}}_z
    \mathrm{d}z'\,
    \alpha_\mathrm{B}
    n_\mathrm{H}(\mathbf{x},z')
    x_\mathrm{e}(\mathbf{x},z')
    \frac{\mathrm{d}t}{\mathrm{d}z'},
    \label{eq:Nrec}
\end{equation}
where $\alpha_\mathrm{B}$ is the case-B recombination \cite{1997MNRAS.292...27H}, $n_\mathrm{H}$ is the number density of hydrogen, and $x_\mathrm{e}$ is the ionized fraction of the cell.

We apply the criterion in Eq.~(\ref{eq:criterion}) by decreasing the filtering scale from $R_\mathrm{max}$ to the cell size. Following \citet{2021MNRAS.502..463T}, we set $R_\mathrm{max}=30\,\mathrm{Mpc}$, which is sufficiently large for the high-redshift range considered in this work. If the criterion is not satisfied even at the cell scale, the cell is treated as partially ionized, with its ionized fraction given by
\begin{equation}
    x_\mathrm{e}(\mathbf{x},z)
    =
    \min\left\{
    \max\left[
    N_\mathrm{ion}(\mathbf{x},z)
    -
    N_\mathrm{rec}(\mathbf{x},z),
    0
    \right],
    1
    \right\}.
    \label{eq:partial_ionized}
\end{equation}

\subsection{escape fraction as a function of popIII star mass and  halo mass}
\label{subsec:escapefraction}

To accurately model the reionization history, it is essential to quantify how efficiently ionizing photons produced by the Pop~III stars escape from their host halos into the IGM. 
The escape fraction, $\mathscr{f}_\mathrm{esc}$, strongly depends on both the mass of the halo and the stellar mass of the central Pop~III star, as these determine the depth of the gravitational potential well and the strength of radiative feedback.  Previous semi-analytic studies have often assumed a constant escape fraction \cite{2020ApJ...897...95V, 2025NatAs...9.1268G}, but radiation–hydrodynamic (RHD) simulations reveal that $\mathscr{f}_\mathrm{esc}$ can vary by orders of magnitude across different mass scales {\cite{2018MNRAS.480.1925T}, and it affects the 21-cm signal not only through ionization but also through photo-heating \cite{2021MNRAS.502..463T}.} In this work, we adopt the fitting relation obtained from one-dimensional RHD simulations presented in \cite{2021MNRAS.502..463T}, which enables us to account for the impact of Pop~III stellar mass on the 21-cm signal through the ionization and heating of the surrounding IGM. This approach provides a physically motivated connection between the properties of Pop~III stars and their ionizing impact on the IGM, distinguishing our model from previous works \cite{2020ApJ...897...95V, 2025NatAs...9.1268G} that employed simplified prescriptions.

In \cite{2021MNRAS.502..463T}, the relationship between the escape fraction of ionizing photons and both halo mass and stellar mass is examined. The escape fraction for a halo with mass $M_\mathrm{h}$ that hosts a Pop~III star with mass $M_\mathrm{s}$ is determined using one-dimensional radiative hydrodynamics (RHD) simulations. The study provides a fitting formula to describe this relationship as 
\begin{equation}
\begin{aligned}
\mathscr{f}_\mathrm{esc}(M_\mathrm{h}, M_\mathrm{s})
&= \max \Bigg[
-18.14\, \left(\frac{M_\mathrm{s}}{M_\odot} \right)^{-0.67}
\left( \frac{M_\mathrm{h}}{10^6\, M_\odot} \right)
\\
&\quad + 0.97,\ 0
\Bigg].
\end{aligned}
\label{eq:fesc}
\end{equation}

\begin{figure}
    \centering
    \includegraphics[width=\linewidth]{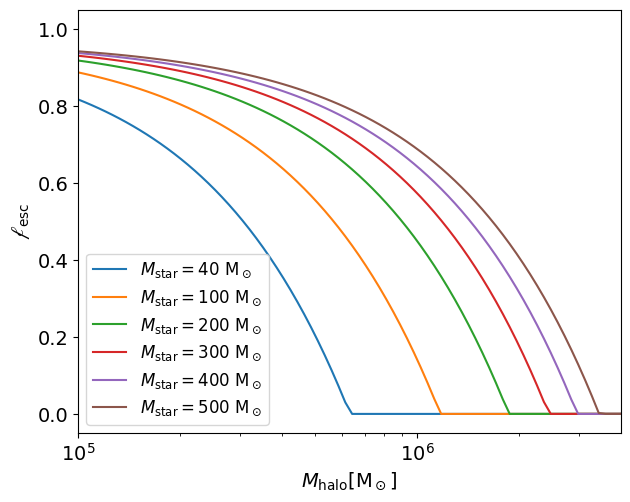}
    \caption{Escape fraction–halo mass relation. Relationship between the ionizing photon escape fraction, $\mathscr{f}_\mathrm{esc}$, and halo mass, $M_\mathrm{h}$, shown for several stellar masses, $M_\mathrm{s}$. $\mathscr{f}_\mathrm{esc}$ decreases with increasing $M_\mathrm{h}$ due to enhanced absorption by hydrogen within more massive haloes, while it increases with increasing $M_\mathrm{s}$ as stronger ionizing emission drives faster expansion of the ionized bubble inside the halo.}
    \label{fig:fesc vs Mhalo}
\end{figure}

Figure \ref{fig:fesc vs Mhalo} illustrates the relationship between $\mathscr{f}_\mathrm{esc}$ and $M_\mathrm{h}$ for various values of $M_\mathrm{s}$. As the halo mass $M_\mathrm{h}$ increases, the escape fraction $\mathscr{f}_\mathrm{esc}$ decreases. This decline occurs because ionizing photons are absorbed by abundant hydrogen within the halo. In contrast, as the star mass $M_\mathrm{s}$ increases, $\mathscr{f}_\mathrm{esc}$ also increases. This increase is due to the rapid expansion of the ionized bubble inside the halo, which is driven by the strong emission from a massive star.

In our simulations, the escape fraction at redshift z is averaged over the halo mass as
\begin{equation}
    f_\mathrm{esc} (z, M_\mathrm{s}) =
    \frac{\int_{M_\mathrm{cool}}^{\infty} \mathrm{d}M_\mathrm{h} \frac{\mathrm{d}n}{\mathrm{d}M_\mathrm{h}} \mathscr{f}_\mathrm{esc}(M_\mathrm{h},M_\mathrm{s})}{\int_{M_\mathrm{cool}}^\infty \mathrm{d}M_\mathrm{h} \frac{\mathrm{d}n}{\mathrm{d}M_\mathrm{h}}}.
    \label{eq:averaged_fesc}
\end{equation}
In this work, we utilize the Sheth-Mo-Tormen mass function \cite{2001MNRAS.323....1S} to describe the mass function $\frac{\mathrm{d}n}{\mathrm{d}M_\mathrm{h}}$ in the equation. The minimum halo mass necessary for gas to cool sufficiently to allow for star formation is referred to as $M_\mathrm{cool}$. It is important to note that $M_\mathrm{cool}$ is influenced by Lyman-Werner feedback. The details regarding this feedback mechanism and the determination of $M_\mathrm{cool}$ will be discussed in the following section.

In applying Eq.~(\ref{eq:averaged_fesc}), we treat $M_\mathrm{s}$ as the characteristic mass of an individual Pop~III star. Specifically, we assume that each occupied halo hosts a single Pop~III star with the same characteristic mass $M_\mathrm{s}$, and we do not impose an explicit stellar-mass--halo-mass relation. This simplifying assumption is motivated by the simulations of \citet{2014ApJ...792...32S}, which found no clear correlation between the mass of the primary Pop~III star and that of its host minihalo. Accordingly, $M_\mathrm{s}$ is treated as a model parameter independent of halo mass throughout this work.

\begin{figure}
    \centering
    \includegraphics[width=1\linewidth]{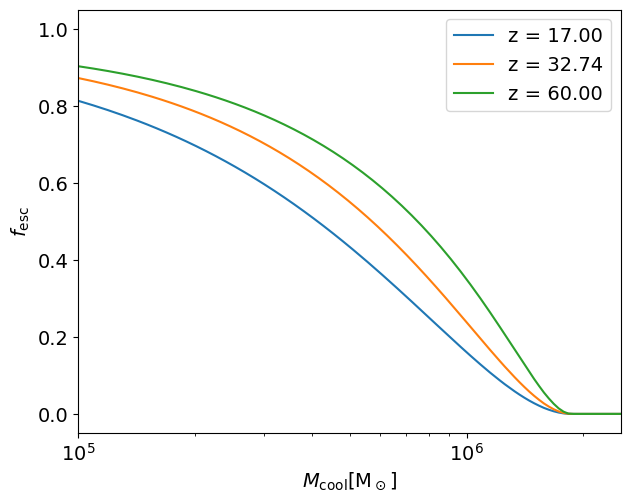}
    \caption{Halo-mass-averaged escape fraction $f_\mathrm{esc}$ as a function of the threshold halo mass for star formation $M_\mathrm{cool}$ in the case of $M_{\rm s}=200 M_\odot$. The escape fraction $f_\mathrm{esc}$ decreases with increasing $M_\mathrm{cool}$ because star formation becomes restricted to more massive haloes that individually exhibit smaller escape fractions, $\mathscr{f}_\mathrm{esc}$. It declines toward lower redshifts, reflecting the increasing contribution of massive haloes that form preferentially at later times.}
    \label{fig:averaged_fesc}
\end{figure}
Figure \ref{fig:averaged_fesc} shows the relationship between the halo-mass-averaged escape fraction $f_\mathrm{esc}$ and $M_\mathrm{cool}$ for the $M_\mathrm{s}=200 \,\mathrm{M_\odot}$ case at various redshifts. As $M_\mathrm{cool}$ increases, $f_\mathrm{esc}$ decreases. This occurs because star formation is primarily limited to more massive halos, which have a smaller individual escape fraction $\mathscr{f}_\mathrm{esc}$. Additionally, as the redshift decreases, $f_\mathrm{esc}$ also decreases. This trend is due to the fact that massive halos tend to form at lower redshifts.

\subsection{Photo-heating of IGM by UV radiation}
Photo-heating by UV photons produces a warm, 21-cm--emitting layer surrounding ionized regions \cite{2014MNRAS.445.3674Y, 2018MNRAS.480.1925T}.
In our simulations, however, this heating structure cannot be explicitly resolved because the characteristic mean free path of the relevant UV photons is shorter than the simulation grid size. 
We therefore model UV heating in a sub-grid manner. Following \cite{2021MNRAS.502..463T}, we subdivide each cell into three components: an ionized region, a cold neutral region, and a heated neutral region. 
Assuming $x_\mathrm{HI}=0$ in the ionized region and $T_\mathrm{S} \gg T_\mathrm{CMB}$ in the heated region, the 21-cm brightness temperature can be expressed as
\begin{equation}
    \delta T_\mathrm{b,ion} = 0 \ \mathrm{mK},
\end{equation}
\begin{equation}
    \delta T_\mathrm{b, cold}= 38.7 (1+\delta) \left(\frac{1+z}{20}\right)^{1/2} \frac{T_\mathrm{S}-T_\mathrm{CMB}(z)}{T_\mathrm{S}} \ \mathrm{mK},
\end{equation}    
\begin{equation}
    \delta T_\mathrm{b, heat}= 38.7 (1+\delta) \left(\frac{1+z}{20}\right)^{1/2} \ \mathrm{mK},
\end{equation}
respectively.
The brightness temperature of the cell is written as a weighted average of the brightness temperature of these three regions:
\begin{equation}
    \delta T_\mathrm{b,grid} = \sum_i f_i \delta T_{\mathrm{b},i} \ \ \ (i=\mathrm{ion},\, \mathrm{cold},\, \mathrm{heat}).
\end{equation}
Here, $f_i$ is the volume fraction of each region. 
In \cite{2021MNRAS.502..463T}, radiative hydrodynamic (RHD) simulations were performed to analyze the ratio of the volume of heated regions to that of ionized regions, defined as $\gamma_\mathrm{h/i} \equiv f_\mathrm{heat}/f_\mathrm{ion}$. They found that the dependence of $\gamma_\mathrm{h/i}$ on halo and stellar masses is negligible, and they also provided a fitting formula for $\gamma_\mathrm{h/i}$ as 
\begin{equation}
    \log(\gamma_\mathrm{h/i}) = -3.11\log(1 + z) + 5.23.
\end{equation}
In the RHD simulations, the boundary between the ionized region and the heated region is defined as the shell where the neutral fraction is $1\%$, while the boundary between the heated region and the cold region is defined as the shell where $\delta T_b$ becomes positive. Under these definitions, the ionized fraction in the heated region is below $10\%$, and the heated region is therefore approximated as fully neutral, $x_\mathrm{HI} = 1$, for simplicity \cite{2021MNRAS.502..463T}.
By combining this relationship with $f_\mathrm{ion}$ obtained in the simulations as shown in equation (\ref{eq:partial_ionized}), along with the normalization condition $\sum_if_i = 1$, we can determine all $f_i$ values for each cell.

\subsection{LW feedback}
\label{subsec:LW_feedback}
LW photons, which have energies ranging from  $11.2$ to $13.6$ \ $\mathrm{eV}$, can dissociate molecular hydrogen. Since molecular hydrogen serves as the primary coolant in primordial gas clouds where Pop~III stars form, its dissociation through LW radiation emitted by the first stars suppresses the formation of subsequent stars. This negative feedback mechanism is incorporated \cite{2021MNRAS.502..463T} and also in this work. 

To quantify this feedback effect, we adopt the framework established by \cite{2014MNRAS.445..107V}, which relates the minimum halo mass required for Pop~III star formation to the local LW background intensity. The minimum halo mass, denoted as \(M_\mathrm{cool}\) in equation (\ref{eq:averaged_fesc}), above which halos can host Pop~III stars, depends on the LW intensity, \(J_\mathrm{LW}\).
The relationship between $M_\mathrm{cool}$ and $J_\mathrm{LW}$ found in \cite{2014MNRAS.445..107V} is given by
\begin{equation}
\begin{split}
    M_\mathrm{cool} 
    &= 2.5 \times 10^5 
       \left( \frac{1+z}{26} \right)^{-1.5} \\
    &\quad \times \!\left[ 1 + 6.96\,(4\pi J_\mathrm{LW}(z))^{0.47} \right]
       M_\odot ,
    \label{eq:Mcool}
\end{split}
\end{equation}
where $J_\mathrm{LW}(z)$ is given in units of $[10^{-21} \ \mathrm{erg \ s^{-1} \ cm^{-2} \ Hz^{-1} \ str^{-1}}]$. The averaged LW intensity, $J_\mathrm{LW}(z)$, at each redshift is calculated by averaging LW intensity at each simulation grid at that redshift, $\mathcal{J}_\mathrm{LW}(\mathbf{x},z)$. $\mathcal{J}_\mathrm{LW}(\mathbf{x},z)$ is calculated by summing the contributions from the emissivity of the sphere centered at $\mathbf{x}$ with a radius of $r_\mathrm{p}$
\begin{equation}
    \mathcal{J}_\mathrm{LW} (\mathbf{x}, z) = \int^{z_\mathrm{max}}_z \mathrm{d}z' \frac{1}{4\pi} \frac{1}{4\pi r_\mathrm{p}^2} \frac{\mathrm{d}\varepsilon(\mathbf{x}, z')}{\mathrm{d}z'},
    \label{eq:LWintensity}
\end{equation}
where $r_\mathrm{p}$ corresponds to the proper light-travel distance between $z'$ and $z$, and $\varepsilon(\mathbf{x}, z)$ is the LW specific emissivity. The emissivity $\varepsilon(\mathbf{x}, z)$ is assumed to be proportional to the growth of the collapsed mass fraction inside halos, $f_\mathrm{coll}$, and can be written as 
\begin{equation}
    \frac{\mathrm{d}\varepsilon(\mathbf{x}, z')}{\mathrm{d}z'} = \left( \frac{N_\mathrm{LW}E_\mathrm{LW}}{\Delta \nu_\mathrm{LW}} \right) f_\ast f_\mathrm{esc,LW} \bar{n}_\mathrm{b,0} (1+\bar{\delta}_R) \frac{\mathrm{d}V}{\mathrm{d}z'} \frac{\mathrm{d}f_\mathrm{coll}}{\mathrm{d}t}~.
    \label{eq:LWemissivity}
\end{equation}
Here, the factor $\frac{N_\mathrm{LW}E_\mathrm{LW}}{\Delta \nu_\mathrm{LW}}$ represents the energy of LW photons per stellar baryon per frequency, and $f_\mathrm{esc,LW}$ denotes the escape fraction of LW photons, which is assumed to be equal to the escape fraction of ionizing photons, denoted as $f_\mathrm{esc}$. The remaining part the RHS of the equation is related to the star formation rate within the shell between $z'$ and $z'+\mathrm{d}z'$. In this context, $f_\ast$ is the star formation efficiency, $\bar{n}_\mathrm{b,0}$ denotes the present baryon number density, $\bar{\delta}_R$ indicates the density fluctuation averaged over the scale $R$, and $\frac{\mathrm{d}V}{\mathrm{d}z'}$ refers to the volume of the shell. 
Following the prescription in the original 21cmFAST implementation\cite{2011MNRAS.411..955M}, the fraction of matter collapsed into halos above the minimum
mass for Pop~III star formation, $M_\mathrm{cool}$, is denoted by
$f_\mathrm{coll}$ and calculated as
\begin{equation}
\begin{aligned}
    f_\mathrm{coll}&(\mathbf{x}, z', R', S_\mathrm{min})\\ =
    &\frac{\bar{f}_\mathrm{ST}(z', S_\mathrm{min})}{\bar{f}_\mathrm{PS,nl}(z', S_\mathrm{min}, R')} \mathrm{erfc}\left[\frac{\delta_c - \delta_\mathrm{nl}^{R'}}{\sqrt{2[S_\mathrm{min}-S^{R'}]}}\right],
\end{aligned}
\end{equation}
where $S_\mathrm{min}$ and $S^{R'}$ are the variances of the smoothed linear density field at halo mass scales $M_\mathrm{cool}$ and $R'$, respectively. $\bar{f}_\mathrm{ST}$ is the mean Sheth-Tormen collapsed fraction and $\bar{f}_\mathrm{PS,nl}$ is the mean Press-Schechter collapsed fraction averaged over scale $R'$.

In our framework, $f_\ast$ is defined as the fraction of the total baryonic mass contained in halos with masses above $M_\mathrm{cool}$ within a simulation cell that is converted into Pop~III stars, rather than as a stellar-to-halo mass conversion factor for each individual halo. We assume that each star-hosting halo hosts a single Pop~III star of mass $M_\mathrm{s}$. Therefore, for fixed $M_\mathrm{s}$ and a fixed halo population, a larger $f_\ast$ corresponds to a larger number, or equivalently a larger occupation fraction, of halos hosting a Pop~III star, rather than to a larger stellar mass in each halo.

Combining Eqs. (\ref{eq:LWintensity}) and (\ref{eq:LWemissivity}), the LW intensity in each grid can be written as 
\begin{equation}
\begin{aligned}
    \mathcal{J}_\mathrm{LW} (\mathbf{x}, z) = & \frac{f_\ast  \bar{n}_\mathrm{b,0}c}{4\pi} \frac{N_\mathrm{LW}E_\mathrm{LW}}{\Delta \nu_\mathrm{LW}} \\ & \times \int^{z_\mathrm{max}}_z \mathrm{d}z' f_\mathrm{esc,LW} (1+z')^3 (1+\bar{\delta}_R) \frac{\mathrm{d}f_\mathrm{coll}}{\mathrm{d}z'}~.
\end{aligned}
\end{equation}

An increase in the LW flux raises $M_\mathrm{cool}$, thereby
reducing $f_\mathrm{coll}$, the fraction of matter collapsed into halos massive enough to host Pop~III star formation.
The resulting suppression of the SFRD decreases the LW flux
$J_\mathrm{LW}$, which in turn lowers $M_\mathrm{cool}$.

To calculate $M_\mathrm{cool}$ and $J_\mathrm{LW}$ consistently, we implement an iteration process.
Initially, we calculate $\mathcal{J}_\mathrm{LW}$ and $M_\mathrm{cool}$ according to equations (\ref{eq:Mcool}) and (\ref{eq:LWintensity}). Next, we determine the temporal values of $f_\mathrm{esc}$ and $f_\mathrm{coll}$. Using these temporal values,  we then recalculate $\mathcal{J}_\mathrm{LW}$ and $M_\mathrm{cool}$. This process is repeated until we obtain the final values of $f_\mathrm{esc}$ and $f_\mathrm{coll}$, which are then used to calculate the ionization field.

\begin{figure}
    \includegraphics[width=\linewidth]{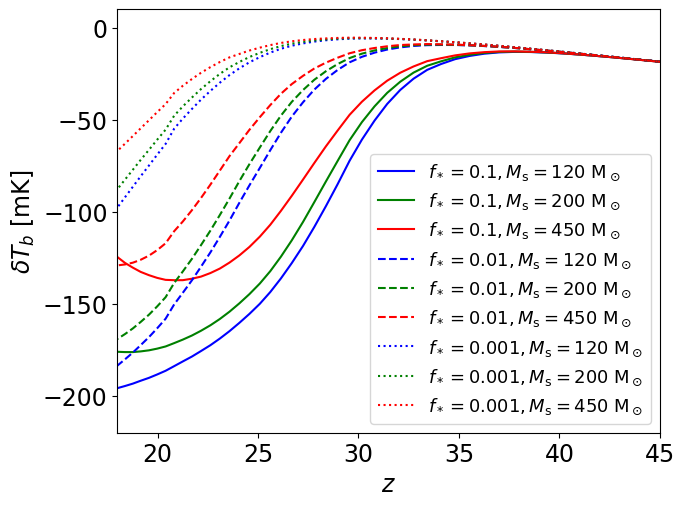}
    \caption{The global 21-cm brightness temperatures as functions of redshift. The solid, dashed and dotted lines are cases of $f_\ast = 0.1, 0.01, 0.001$ respectively. The red, green, blue lines are cases of $M_\mathrm{s} = 450, 200, 120 \,\mathrm{M_\odot}$ respectively.}
    \label{fig:Tb}
\end{figure}

Figure \ref{fig:Tb} shows the calculated 21-cm global signal for several values of star formation efficiency $f_\ast$ and the star mass $M_\mathrm{s}$. Here, $M_\mathrm{s}$ specifies the mass of an individual Pop~III star, whereas $f_\ast$ controls the abundance of star-hosting halos within each simulation cell. Since Pop~II stars emerge as the dominant sources of photons at lower redshifts, we compute the 21-cm signal only for $z>18$. An absorption feature appears at $z<40$. A higher $f_\ast$ results in a deeper absorption, whereas a larger $M_\mathrm{s}$ leads to a shallower one. In the following, we explain the physical origin of these trends based on the evolution of the star formation rate density (SFRD), radiative feedback, and ionization state.

\begin{figure}
    \centering
    \includegraphics[width=\linewidth]{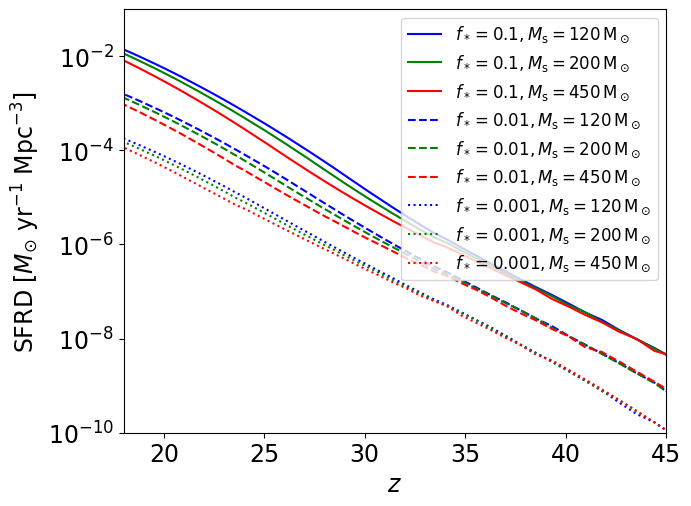}
    \caption{The star formation rate density as a function of redshift. The line styles and colors correspond to the same parameter values as Figure \ref{fig:Tb}.}
    \label{fig:sfrd}
\end{figure}

Figure \ref{fig:sfrd} shows the evolution of the SFRD. A larger $f_\ast$ leads to a large SFRD. This is because star formation occurs in a larger number of halos. As a result, more Ly$\alpha$ photons are produced, which enhances the Wouthuysen-Field coupling. This drives the spin temperature $T_S$ closer to the gas temperature $T_K$. Since $T_K<T_\mathrm{CMB}$, a larger $f_\ast$ leads to a deeper absorption feature in the global 21-cm signal.

\begin{figure}
    \centering
    \includegraphics[width=\linewidth]{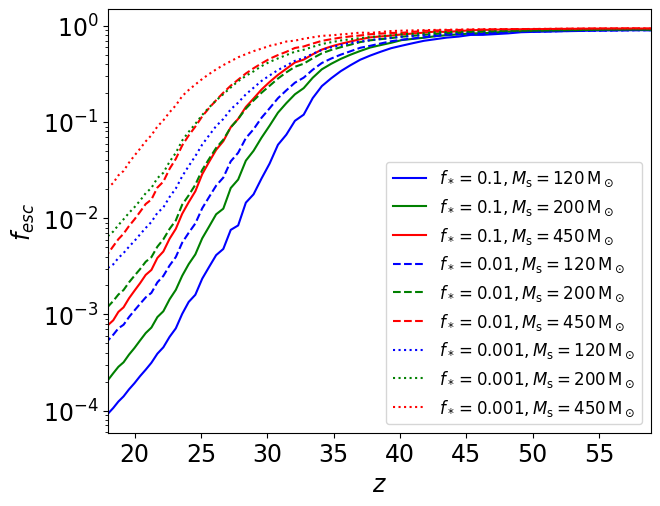}
    \caption{The halo-mass-averaged escape fraction given by eq. \ref{eq:averaged_fesc} as a function of redshift. The line styles and colors correspond to the same parameter values as Figure \ref{fig:Tb}.}
    \label{fig:fesc}
\end{figure}

\begin{figure}
    \centering
    \includegraphics[width=\linewidth]{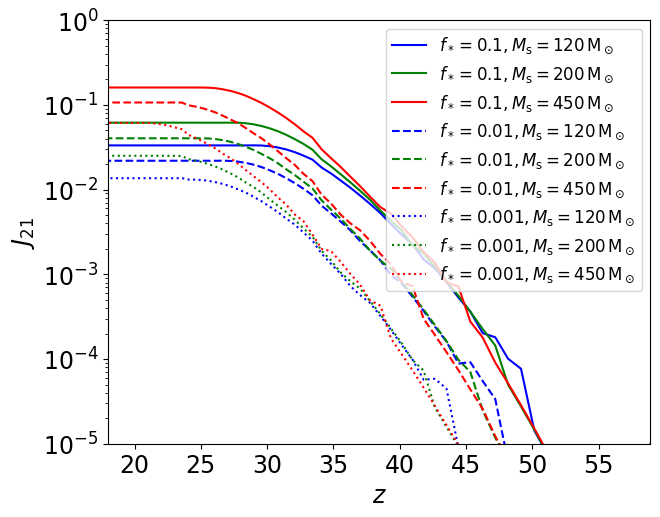}
    \caption{The averaged LW intensity, $J_\mathrm{LW}(z)$, as a function of redshift. The line styles and colors correspond to the same parameter values as Figure \ref{fig:Tb}.}
    \label{fig:J21}
\end{figure}

\begin{figure}
    \centering
    \includegraphics[width=\linewidth]{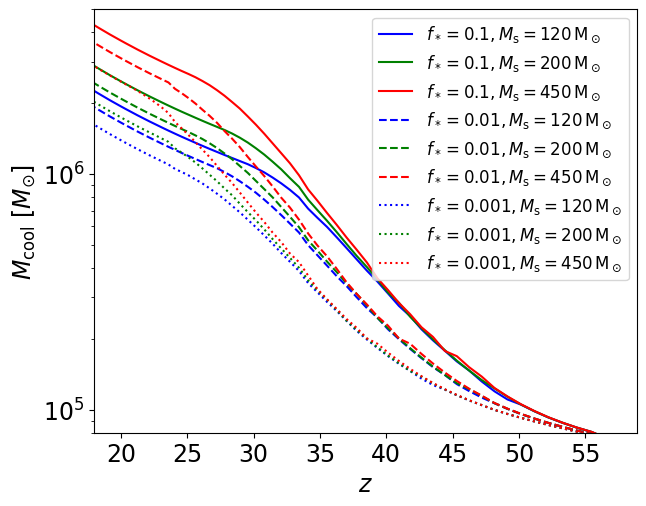}
    \caption{The minimum halo mass, $M_\mathrm{cool}$, as a function of redshift. The line styles and colors correspond to the same parameter values as Figure \ref{fig:Tb}.}
    \label{fig:Mcool}
\end{figure}

By contrast, the dependence on $M_\mathrm{s}$ is primarily governed by radiative feedback from Pop~III stars. As shown in Figure \ref{fig:fesc}, a larger $M_\mathrm{s}$ leads to a higher halo-mass-averaged escape fraction. This enhances the LW intensity, as shown in Fig.~\ref{fig:J21}, and raises the minimum halo mass $M_\mathrm{cool}$, as shown in Fig.~\ref{fig:Mcool}, thereby suppressing star formation in low-mass halos. As a result, the SFRD becomes smaller for a larger $M_\mathrm{s}$, leading to fewer Ly$\alpha$ photons and weaker WF coupling.

\begin{figure}
    \centering
    \includegraphics[width=1\linewidth]{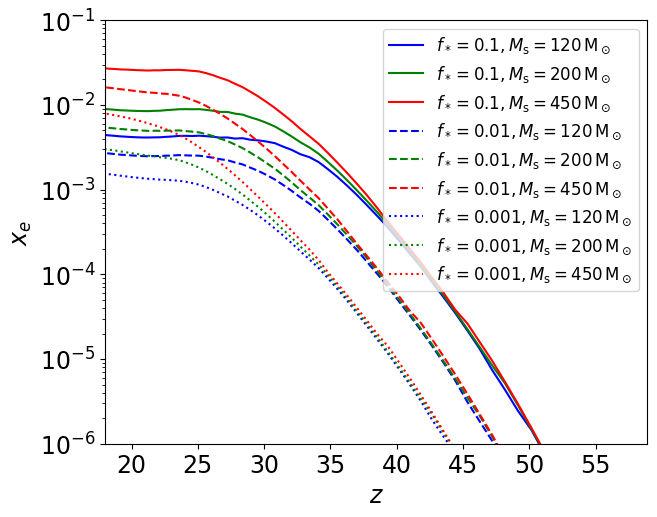}
    \caption{The ionized fraction as a function of redshift. The line styles and colors correspond to the same parameter values as Figure \ref{fig:Tb}.}
    \label{fig:xe}
\end{figure}

In addition to the suppression WF coupling, a larger $M_\mathrm{s}$ also enhances the ionization heating of the IGM. As shown in Fig.~\ref{fig:xe}, the ionized fraction becomes higher for a larger $M_\mathrm{s}$, due to a larger escape fraction. This indicates stronger heating. As a result, the gas temperature $T_K$ approaches the CMB temperature $T_\mathrm{CMB}$, which further makes the absorption feature in the global 21-cm signal shallower.

\section{Fisher forecast}
\label{sec:fisher forecast}

To assess how effectively observations of the 21-cm global signal can constrain the properties of Pop~III stars, we utilize the Fisher matrix method, following~\citet{2010PhRvD..82b3006P}. The Fisher matrix is defined as
\begin{equation}
\mathbf{F}_{i j} \equiv-\left\langle\frac{\partial^2 \ln (\mathcal{L})}{\partial p_i \partial p_j}\right\rangle,
\end{equation}
where $\mathcal{L}$ is the likelihood function and $p_i$ denotes parameter to be constrained. This Fisher matrix is also written as
\begin{equation}
\mathbf{F}_{i j} = \frac{1}{2} \mathrm{Tr} \left[C^{-1}C_{,i}C^{-1}C_{,j} + C^{-1}(\mu_{,i}\mu_{,j}^T + \mu_{,j}\mu_{,i}^T )\right],
\end{equation}
where $C$ is the covariance matrix defined below and $\mu$ represents the expected values of the observable. For 21-cm signal observations, the key observable is the sky temperature $T_\mathrm{sky}$, which is given by the sum of the foreground emission and the cosmological 21-cm signal as
\begin{equation}
T_\mathrm{sky} = T_\mathrm{fg} + \delta T_b.
\end{equation}

To forecast constraints on the Pop~III parameters in the presence of foreground emission, we follow \citet{2010PhRvD..82b3006P} and model the foreground temperature as an $N$th-order log-polynomial,
\begin{equation}
    \ln T_{\rm fg}(\nu)
    =
    \sum_{i=0}^{N}
    a_i
    \left[
    \ln\left(\frac{\nu}{\nu_{\rm ref}}\right)
    \right]^i .
    \label{eq:foreground_fit}
\end{equation}

Similar smooth parametric foreground models have also been used in analyses of global-signal observations. 
The EDGES Low-Band analysis employed a different five-term smooth basis \citep{2018Natur.555...67B}, while the SARAS~3 analysis examined several orders of a log-polynomial foreground model \citep{2022NatAs...6..607S}.
In both analyses, the smooth foreground and the 21-cm signal were inferred simultaneously.

The coefficients $a_i$ are included as nuisance parameters and varied jointly with the Pop~III parameters in the Fisher analysis.  The fiducial values of $a_i$ are determined by fitting the log-polynomial model to the sky-averaged GSM2008 temperature spectrum \citep{2008MNRAS.388..247D}. For both foreground fitting and Fisher analysis, we adopt the frequency range $45.8 \leq \nu \leq 74.7\,\mathrm{MHz}$, corresponding to the redshift range $18 \leq z \leq 30$. This lower-redshift boundary is adopted to avoid the regime where the Pop~II contribution is expected to become important, while the upper-redshift boundary corresponds to the low-frequency limit, below which ground-based observations become increasingly challenging. The reference frequency $\nu_{\rm ref}$ is fixed to the center of this band.

\begin{figure}
    \centering
    \includegraphics[width=\linewidth]{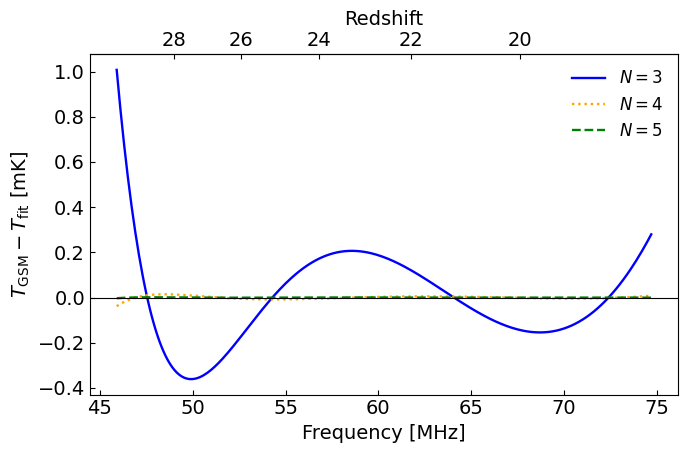}
    \caption{Temperature residuals between the sky-averaged GSM2008 foreground spectrum and the best-fitting log-polynomial models, $T_{\rm GSM}-T_{\rm fit}$. The solid, dotted, and dashed curves correspond to polynomial orders $N=3$, $4$, and $5$, respectively. The models are obtained by fitting $\ln T_{\rm GSM}$ with an $N$th-order polynomial in $\ln(\nu/\nu_{\rm ref})$, where $\nu_{\rm ref}$ is fixed at the center of the fitting band. The lower and upper horizontal axes show the observing frequency and the corresponding redshift, respectively.}
    \label{fig:foreground_residual}
\end{figure}

Figure~\ref{fig:foreground_residual} shows the residuals between the sky-averaged GSM2008 spectrum and the best-fitting log-polynomial foreground models. The residual decreases as the polynomial order is increased. We therefore consider $N=3$--$5$ in the Fisher analysis to quantify the sensitivity of the forecast to the foreground model.

We include thermal noise in the covariance matrix and assume that the noise is uncorrelated between frequency channels. The covariance matrix is therefore diagonal,
\begin{equation}
C_{nm} = \sigma_n^2 \delta_{nm},
\label{eq:covariance}
\end{equation}
where
\begin{equation}
\sigma_n^2 = \frac{T^2_\mathrm{sky}(\nu_n)}{t_\mathrm{int}B} .
\label{eq:error}
\end{equation}
Here, $t_\mathrm{int}$ is the integration time and $B$ is the channel bandwidth. We assume $t_\mathrm{int} = 1000\, \mathrm{hr}$ and $B = 0.1 \ \mathrm{MHz}$.
Under these assumptions, the Fisher matrix is given by
\begin{equation}
\begin{aligned}
    F_{ij}
    =
    \sum_{n=1}^{N_\mathrm{channel}}
    \left(2 + B t_\mathrm{int}\right)
    \frac{\partial \ln T_\mathrm{sky}(\nu_n)}
         {\partial p_i}
    \frac{\partial \ln T_\mathrm{sky}(\nu_n)}
         {\partial p_j}.
\end{aligned}
\label{eq:fisher}
\end{equation}
Here, the parameter set $p_i$ is $\boldsymbol{p}=\left(a_0,a_1,\ldots,a_N,f_\ast,M_\mathrm{s}\right)$, containing both the foreground coefficients and the Pop~III parameters. Following the formulation of \citet{2010PhRvD..82b3006P}, the prefactor “2+” arises because both the mean signal and the covariance depend on the astrophysical and foreground parameters. The first term represents the information contribution from the parameter dependence of the noise covariance, i.e., the change in the overall variance amplitude with respect to the model parameters. The second term corresponds to the usual Fisher information from the derivative of the mean signal, reflecting how the global 21-cm spectrum itself varies with astrophysical parameters. See Appendix A for the derivation of the Fisher matrix.

The covariance matrix of the model parameters is estimated from the inverse Fisher matrix,
\begin{equation}
    \Sigma_{ij}
    =
    \left(F^{-1}\right)_{ij}.
\end{equation}

\begin{figure*}
    \centering
    \includegraphics[width=\linewidth]{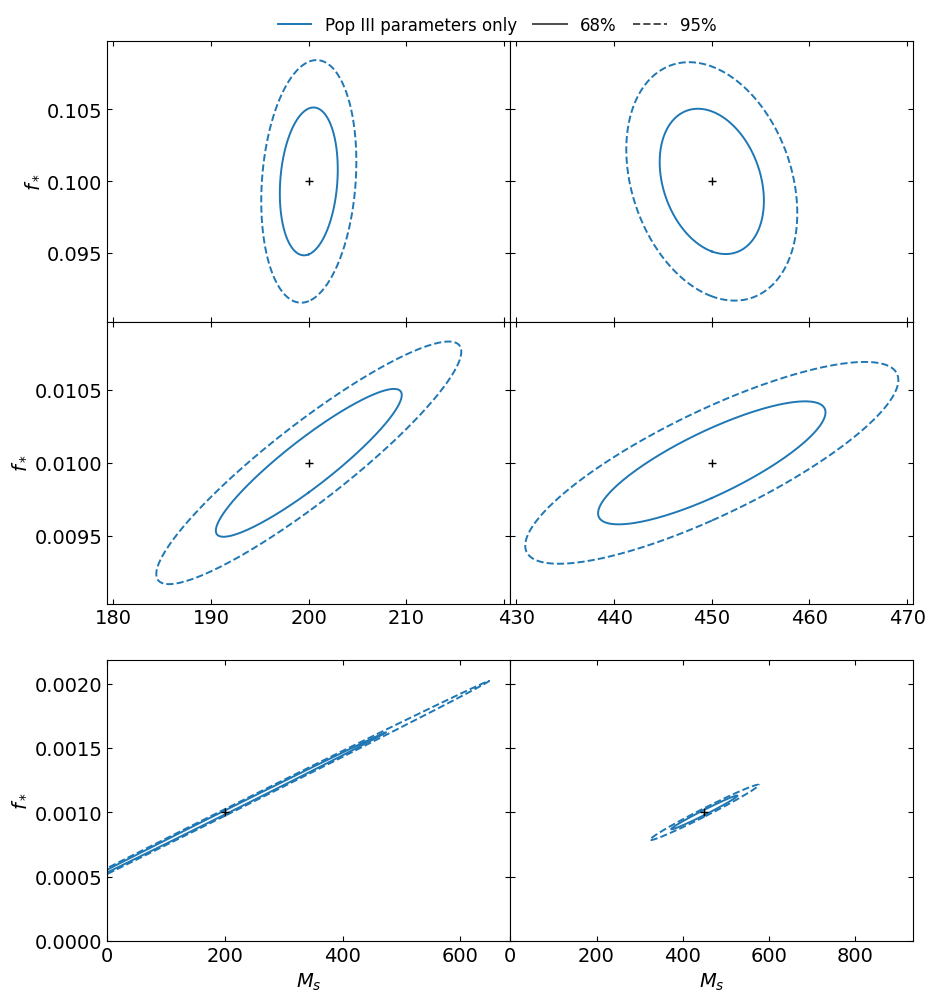}
    \caption{Forecasted constraints on the Pop~III star-formation efficiency, $f_\ast$, and stellar mass, $M_\mathrm{s}$, with the foreground parameters fixed at their fiducial values, while the foreground temperature is included in the fiducial sky temperature. The rows correspond to $f_\ast=0.1$, $0.01$, and $0.001$ from top to bottom, and the left and right columns correspond to $M_\mathrm{s}=200$ and $450\,M_\odot$, respectively. The solid and dashed contours show the 68\% and 95\% confidence regions, and the black crosses indicate the fiducial parameter values. An integration time of $1000\,\mathrm{h}$ is assumed. The upper two rows share the same horizontal scale, whereas a wider $M_\mathrm{s}$ range is used for the bottom row.}
    \label{fig:fisher_PopIIIonly}
\end{figure*}

\begin{figure*}
    \centering
    \includegraphics[width=\linewidth]{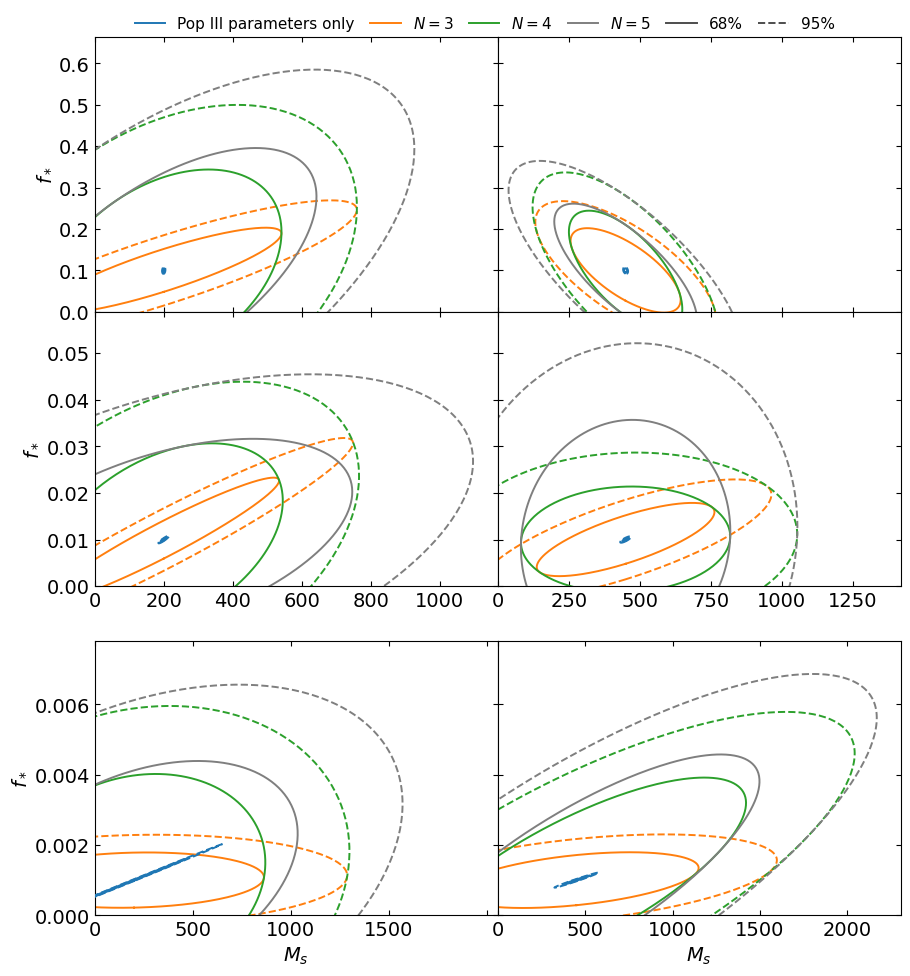}
    \caption{Forecasted constraints on the Pop~III parameters obtained by simultaneously including the foreground coefficients in the Fisher analysis. The foreground is modeled using the log-polynomial form given in Eq.~(\ref{eq:foreground_fit}), with the reference frequency $\nu_\mathrm{ref}$ fixed. The blue contours show the results with the foreground parameters fixed at their fiducial values, while the orange, green, and gray contours show the constraints after marginalizing over the foreground coefficients for $N=3$, $4$, and $5$, respectively. These models contain four, five, and six foreground coefficients. The solid and dashed contours denote the 68\% and 95\% confidence regions. The rows correspond to $f_\ast=0.1$, $0.01$, and $0.001$ from top to bottom, and the left and right columns correspond to $M_\mathrm{s}=200$ and $450\,M_\odot$, respectively. An integration time of $1000\,\mathrm{h}$ is assumed.}
    \label{fig:fisher_foreground}
\end{figure*}

Figure~\ref{fig:fisher_PopIIIonly} shows the forecasted constraints when the foreground parameters are fixed at their fiducial values. For $f_\ast=0.1$ and $0.01$, both $f_\ast$ and $M_\mathrm{s}$ can be constrained relatively well, although the degeneracy direction depends on the fiducial Pop~III parameters. For example, at $f_\ast=0.01$ and $M_\mathrm{s}=200\,M_\odot$, the marginalized $1\sigma$ uncertainties are $\sigma(f_\ast)=3.3\times10^{-4}$ and $\sigma(M_\mathrm{s})=6.3\,M_\odot$, corresponding to fractional uncertainties of approximately $3\%$ for both parameters. In contrast, the constraints become substantially weaker for $f_\ast=0.001$, particularly for $M_\mathrm{s}$, because the global signal becomes less sensitive to changes in the Pop~III stellar mass within the frequency range considered.

Figure~\ref{fig:fisher_foreground} shows how the constraints change when the foreground coefficients are simultaneously included as nuisance parameters. Marginalizing over the foreground coefficients substantially broadens the Pop~III confidence regions for all fiducial parameter values. For $f_\ast=0.01$ and $M_\mathrm{s}=200\,M_\odot$, even the $N=3$ foreground model increases the marginalized $1\sigma$ uncertainties to $\sigma(f_\ast)=8.8\times10^{-3}$ and $\sigma(M_\mathrm{s})=2.2\times10^{2}\,M_\odot$, corresponding to approximately $88\%$ and $110\%$ of their fiducial values, respectively. The constraints generally become weaker as the polynomial order is increased from $N=3$ to $N=5$. This degradation occurs because the additional smooth foreground modes overlap with spectral variations induced by the Pop~III parameters, reducing the independent information available to constrain $f_\ast$ and $M_\mathrm{s}$. The foreground marginalization also changes the orientation of the confidence regions, indicating that the apparent degeneracy between the two Pop~III parameters is affected by their correlations with the foreground coefficients.

\section{Discussion}
\label{sec:discussion}

In this section, we discuss the physical interpretation of the trends found in the Fisher analysis. We first consider the constraints obtained with the foreground parameters fixed, in order to isolate the effects of the two Pop~III parameters and clarify the physical origin of their degeneracy. We then discuss how the constraints change when the foreground coefficients are varied simultaneously with the Pop~III parameters. As shown in Fig.~\ref{fig:fisher_PopIIIonly}, the constraints become tighter for larger $f_\ast$.
This is likely because a smaller $f_\ast$ results in a lower redshift at which the absorption feature appears, leading to weak parameter dependence within the range of $z=18$--$30$ we consider.

\begin{figure}
    \centering
    \includegraphics[width=\linewidth]{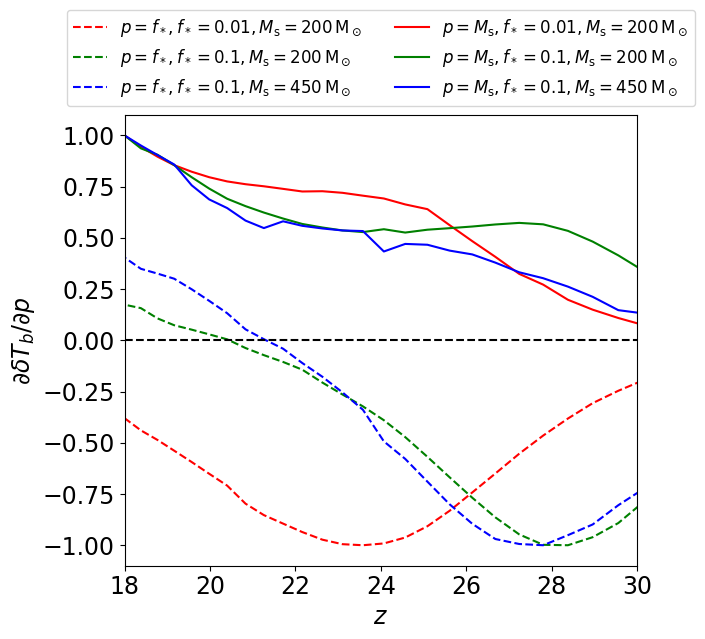}
    \caption{The parameter derivatives of the global 21-cm brightness temperature as functions of redshift, normalized by its maximum absolute value. The dashed and solid curves show the derivatives with respect to $f_\ast$ and $M_\mathrm{s}$ respectively. The green and red curves show the derivatives around $f_\ast = 0.1$ and $f_\ast = 0.01$ respectively; in all cases $M_\mathrm{s} = 200 \ \mathrm{M_\odot}$. The blue curves show those around $f_\ast = 0.1$ and $M_\mathrm{s}=450 \ \mathrm{M_\odot}$.}
    \label{fig:diff_params}
\end{figure}

The confidence ellipses for $f_\ast=0.01$ and $M_\mathrm{s}=200 \ \mathrm{M_\odot}$ exhibit a positive degeneracy between $f_\ast$ and $M_\mathrm{s}$. This can be understood from how these parameters affect the global signal. Immediately after the absorption trough begins to form, a larger $f_\ast$ leads to a deeper trough, while a larger $M_\mathrm{s}$ leads to a shallower trough. As a result, when both parameters increase simultaneously, their effects on the signal partially cancel each other, suppressing the change in the signal. This leads to a positive degeneracy between $f_\ast$ and $M_\mathrm{s}$. This interpretation is consistent with the derivatives $\frac{\partial \delta T_b}{\partial f_\ast}$ and $\frac{\partial \delta T_b}{\partial M_\mathrm{s}}$ shown in Fig.~\ref{fig:diff_params}, which have opposite signs for most of the considered redshift range. From Eq.~(\ref{eq:fisher}), the opposite signs of these derivatives make the off-diagonal element of the Fisher matrix negative. Since the covariance matrix is the inverse of the Fisher matrix, its off-diagonal term encodes the parameter degeneracy. When this term is positive, an increase in one parameter can be compensated by an increase in the other, producing an error ellipse along a positively tilted direction as shown in Fig.\ref{fig:fisher_PopIIIonly}.

In contrast, for $f_\ast = 0.1$, the degeneracy becomes weaker, as shown in the top left panel of Fig.~\ref{fig:fisher_PopIIIonly}. In this large-$f_\ast$ case, the ionization of the IGM proceeds earlier, making heating by UV photons important. Once heating begins to affect the signal, a larger $f_\ast$ enhances the heating and thus makes the absorption trough shallower (see Fig.~\ref{fig:Tb}). As a result, the derivative with respect to $f_\ast$ changes sign from negative to positive at a lower redshift, which is shown as the green dashed line in Fig.~\ref{fig:diff_params}.

By contrast, the effect of $M_\mathrm{s}$ remains qualitatively similar throughout the evolution. Before heating becomes important, a larger $M_\mathrm{s}$, corresponding to a larger escape fraction, strengthens the LW feedback, suppresses early star formation, and weakens the Ly-$\alpha$ coupling, thereby making the absorption trough shallower. After heating becomes important, the stronger ionizing radiation associated with larger $M_\mathrm{s}$ leads to more efficient heating, which again makes the trough shallower. Therefore, the derivative with respect to $M_\mathrm{s}$ remains positive over the relevant redshift range (solid lines in Fig.~\ref{fig:diff_params}).
Since the response to $f_\ast$ changes sign, whereas that to $M_\mathrm{s}$ remains positive, the two parameters no longer affect the signal in the same way for the case of $f_\ast=0.1$. Consequently, the degeneracy between them is reduced. 

The top right panel of Figure~\ref{fig:fisher_PopIIIonly} summarizes the expected constraints for the case where $M_{\rm s}=450\,\mathrm{M}_\odot$ and $f_\ast = 0.1$. As $M_\mathrm{s}$ is increased from $200$ to $450\,\mathrm{M}_\odot$, the heating effect becomes even stronger and the onset of heating shifts to higher redshift (blue dashed line in Fig.~\ref{fig:diff_params}). As a result, the redshift range over which the parameter dependences of the global signal on $f_\ast$ and $M_\mathrm{s}$ become similar increases, and the degeneracy between the two parameters eventually becomes negative.

We next examine how these constraints are affected when the foreground coefficients are included as nuisance parameters.
As shown in Fig.~\ref{fig:fisher_foreground}, marginalizing over the foreground coefficients substantially broadens the confidence contours for all the fiducial cases considered here.
The constraints generally become weaker as the polynomial order $N$ is increased, and the orientation of the degeneracy also changes depending on the fiducial Pop~III parameters.
The constraints obtained with the foreground parameters fixed should therefore be regarded as an idealized reference for assessing the information contained in the global signal before accounting for degeneracies with the foreground.

The degradation of the constraints arises because both the foreground basis functions and the variations of the global 21-cm signal with the Pop~III parameters are smooth functions of frequency.
Although the fiducial foreground coefficients are determined by fitting only the GSM foreground spectrum, the foreground coefficients and the Pop~III parameters are varied simultaneously in the Fisher
analysis.
Consequently, components of the Pop~III parameter derivatives that can be represented by the foreground basis functions become degenerate with variations of the foreground coefficients.
Increasing $N$ introduces additional smooth spectral modes and thus reduces the information available for constraining $f_\ast$ and $M_\mathrm{s}$.

In this study, we conducted the Fisher analysis using simulation results at redshifts $z \geq 18$ to avoid the regime where the Pop~II contribution becomes dominant.
However, the timing of the transition from Pop~III to Pop~II remains uncertain \citep[e.g.][]{2022ApJ...936...45H, 2019ApJ...871..206S} and may depend on $f_\ast$.
A smaller $f_\ast$ delays metal enrichment, allowing Pop~III star formation to persist to lower redshifts. Conversely, for larger $f_\ast$, metal enrichment proceeds more rapidly, and the transition to Pop~II may occur at higher redshifts. Therefore, the present constraints may be conservative when $f_\ast$ is small, whereas they may be optimistic if $f_\ast$ is large.

Moreover, because the transition is expected to be gradual,
a fixed redshift boundary cannot completely separate the Pop~III
and Pop~II contributions. Since the present Fisher analysis does not include the Pop~II
contribution or its associated parameters, it does not account for
the resulting degeneracies with the Pop~III and foreground parameters,
which could further weaken the forecasted constraints.

We assumed that all Pop~III stars have a single mass; however, the
initial mass function typically has a distribution.
Even with the simulation methodology used in this work, it is possible to incorporate an
IMF distribution by assuming a specific IMF and computing an IMF-weighted average of
the escape fraction of ionizing photons, which depends on Pop~III stellar mass, denoted as 
$f_{\mathrm{esc}}(z,M_\mathrm{s})$.
Even though it is beyond the scope of the present study, this approach may enable forecasts regarding constraints on IMF parameters such as the slope of the IMF.

In a previous study investigating the possibility of constraining the Pop~III IMF using the 21\,cm signal, \citet{2025NatAs...9.1268G} demonstrated, through simulations that include the IMF dependence of IGM heating by X-ray binaries formed from Pop~III stars, that the IMF shape can indeed be constrained.
Specifically, assuming global-signal observations with $25~\mathrm{mK}$ noise, they found that an IMF peaking around $200\,M_\odot$ can be distinguished from one peaking around $500\,M_\odot$ at the $3\sigma$ level.
Under the idealized assumption that the foreground can be perfectly removed, our results similarly indicate that the global 21-cm signal contains information about the characteristic mass of Pop~III stars. However, a direct quantitative comparison is difficult because \citet{2025NatAs...9.1268G} considered the IMF dependence of X-ray heating, whereas our study considers heating by UV photons and the stellar-mass dependence of the escape fraction of ionizing photons.
These two approaches therefore include complementary physical effects. Future simulations incorporating the IMF dependence of X-ray heating and UV heating would provide a more complete assessment of the impact of the Pop~III mass on the 21\,cm signal.
Since UV heating affects heating on small scales, whereas X-ray heating affects heating on large scales, it is crucial to include both the heating effects especially when performing analyses based on the power spectrum. Including both UV and X-ray heating may have a significant impact on the precision of IMF constraints and on parameter degeneracies.

As a promising future prospect, an analysis using the power spectrum shows potential.
In this study, we found that the $f_\ast$--$M_\mathrm{s}$ degeneracy can be reduced by observing the global signal before and after the UV heating becomes effective. However, if the spatial scales that $f_\ast$ and $M_\mathrm{s}$ affect the 21\,cm signal differ, the degeneracy may still be broken even with observations taken over a more limited redshift range.
In this study, the simulations assumed that the LW flux was spatially uniform; however, this uniformity may not reflect reality, as spatial variations in the LW flux could influence the signal—especially in analyses based on the power spectrum. 

\section{summary}
\label{sec:summary}
In this study, we investigated the future constraints on the Pop~III star-formation efficiency, denoted as $f_\ast$, and the characteristic stellar mass $M_\mathrm{s}$ using the 21-cm global signal.
First, we conducted simulations that included a detailed model for the Pop~III mass and halo-mass dependence of the escape fraction of ionizing photons, as well as the heating structures formed around halos by UV radiation, and examined the parameter dependence of the 21-cm global signal.
Next, we utilized the results of these simulations to estimate how well future observations of the 21-cm global signal could constrain the Pop~III parameters. This estimation was performed using a Fisher analysis, taking into account thermal noise and assuming an integration time of $1000\,\mathrm{h}$.
Assuming that the foreground spectrum can be perfectly removed in the data analysis, the frequency dependence of the global 21-cm signal provides information for constraining $f_\ast$ and $M_\mathrm{s}$.
Furthermore, when $f_\ast=0.1$, the degeneracy between $f_\ast$ and $M_\mathrm{s}$ decreases.
This reduction in degeneracy is attributed to the effects of UV heating. Before UV heating becomes effective, $f_\ast$ and $M_\mathrm{s}$ influence the absorption trough depth in the opposite direction, whereas once UV heating becomes effective, both parameters make the absorption trough shallower. Thus, observing both epochs helps distinguish the effects of $f_\ast$ and $M_\mathrm{s}$ and reduces their intrinsic degeneracy.
However, when the smooth foreground spectrum must be inferred and removed from the same observational data, its spectral variation becomes strongly degenerate with the changes in the global signal produced by the Pop~III parameters.
Consequently, the constraints on $f_\ast$ and $M_\mathrm{s}$ are substantially weakened, with more flexible foreground descriptions generally producing weaker constraints.
In this work, we examined the constraining power from the global signal over $z=18$--$30$ for all parameter values. We chose this range because, at lower redshifts, the contributions from Pop~II stars and the first galaxies become more significant.
However, since metal enrichment is expected to proceed more rapidly for larger $f_\ast$, the redshift range that can be used to constrain Pop~III parameters may vary depending on these parameter values.
Moreover, because the Pop~III--Pop~II transition is expected to be gradual, a residual Pop~II contribution could introduce additional parameter degeneracies that are not included in the present forecast.
Overall, these results demonstrate that accurate foreground modeling and removal are essential for extracting information about Pop~III properties from observations of the global 21-cm signal.

\begin{acknowledgments}
We are grateful to Seiya Imoto for his work on improving and debugging the simulation code used in this work. This work is supported in part by the JSPS grant numbers 21H04467, 24K00625, and JST FOREST Program JPMJFR20352935 (KI). HS is supported by Yunnan Provincial Key Laboratory of Survey Science with project No. 202449CE340002, the National SKA Program of China (No.2020SKA0110401), NSFC (Grant No.~12103044), and ``High-End Foreign Expert of the Yunnan Revitalization Talents Support Plan of Yunnan Province''.
\end{acknowledgments}

\appendix
\section{Derivation of the Fisher matrix expression}
\label{app:fisher}

In this Appendix, we derive the Fisher matrix formula used in Section~\ref{sec:fisher forecast}, following the approach of \citet{2010PhRvD..82b3006P}.  
For a Gaussian likelihood, the Fisher matrix is given by
\begin{equation}
\label{eq:app_general}
F_{ij} = \frac{1}{2}\mathrm{Tr}\!\left[
C^{-1}C_{,i}C^{-1}C_{,j}
+ C^{-1}(\mu_{,i}\mu_{,j}^{T}+\mu_{,j}\mu_{,i}^{T})
\right],
\end{equation}
where $C$ is the covariance matrix, $\mu$ is the expectation value of the observable, and commas denote derivatives with respect to the parameters $p_i$ and $p_j$.

\vspace{0.5em}
\noindent{\bf (1) Model assumptions.}  
For global 21-cm observations, the observable is the sky-averaged brightness temperature at each frequency channel $\nu_n$,  
\[
\mu_n = T_{\rm sky}(\nu_n) = T_{\rm fg}(\nu_n) + \delta T_b(\nu_n).
\]
The measurement covariance is assumed to be diagonal,
\begin{equation}
C_{nm} = \delta_{nm}\,\sigma_n^2, \qquad
\sigma_n^2 = \kappa\,T_{\rm sky}^2(\nu_n),
\end{equation}
where $\kappa$ is the frequency-independent noise factor
\begin{equation}
\kappa = 
\frac{1}{t_{\rm int}B},
\end{equation}
with $t_{\rm int}$ and $B$ denoting integration time and bandwidth, respectively.  
We neglect any parameter dependence of $\kappa$, so that only $T_{\rm sky}$ depends on the astrophysical and foreground parameters $p_i$.

\vspace{0.5em}
\noindent{\bf (2) Covariance term.}  
For a diagonal covariance matrix, the first term of Eq.~(\ref{eq:app_general}) becomes
\begin{equation}
\frac{1}{2}\mathrm{Tr}(C^{-1}C_{,i}C^{-1}C_{,j})
  =\frac{1}{2}\sum_n
  \left(\frac{1}{\sigma_n^2}\frac{\partial\sigma_n^2}{\partial p_i}\right)
  \left(\frac{1}{\sigma_n^2}\frac{\partial\sigma_n^2}{\partial p_j}\right).
\end{equation}
Since $\sigma_n^2=\kappa T_{\rm sky}^2$, we have
\begin{equation}
\frac{1}{\sigma_n^2}\frac{\partial\sigma_n^2}{\partial p_i}
   = 2\,\frac{\partial\ln T_{\rm sky}}{\partial p_i},
\end{equation}
and hence
\begin{equation}
\frac{1}{2}\mathrm{Tr}(C^{-1}C_{,i}C^{-1}C_{,j})
 = \sum_n
   2\,\frac{\partial\ln T_{\rm sky}}{\partial p_i}
      \frac{\partial\ln T_{\rm sky}}{\partial p_j}.
\end{equation}
This “2” term originates from the parameter dependence of the covariance amplitude---that is, the fact that the noise variance itself scales with $T_{\rm sky}^2$.

\vspace{0.5em}
\noindent{\bf (3) Mean-derivative term.}  
The second term of Eq.~(\ref{eq:app_general}) reduces to
\begin{align}
\frac{1}{2}\mathrm{Tr}\!\left[
C^{-1}(\mu_{,i}\mu_{,j}^{T}+\mu_{,j}\mu_{,i}^{T})
\right]
&= \sum_n
 \frac{1}{\sigma_n^2}
 \frac{\partial T_{\rm sky}}{\partial p_i}
 \frac{\partial T_{\rm sky}}{\partial p_j} \nonumber \\
&= \sum_n
 \frac{1}{\kappa}
 \frac{\partial\ln T_{\rm sky}}{\partial p_i}
 \frac{\partial\ln T_{\rm sky}}{\partial p_j}.
\end{align}
This corresponds to the standard Fisher information from the derivative of the mean signal, representing how the global 21-cm spectrum changes with the parameters.

\vspace{0.5em}
\noindent{\bf (4) Final expression.}  
Combining the two contributions, the Fisher matrix becomes
\begin{equation}
F_{ij}
=
\sum_{n=1}^{N_{\rm ch}}
\left(2+ t_{\rm int}B\right)
\frac{\partial\ln T_{\rm sky}(\nu_n)}{\partial p_i}
\frac{\partial\ln T_{\rm sky}(\nu_n)}{\partial p_j}.
\label{eq:app_final}
\end{equation}
The prefactor “2+” arises because both the covariance and the mean depend on the parameters:  
the “2” term encodes the information from the parameter dependence of the covariance (noise amplitude), while the “$t_{\rm int}B$” term corresponds to the response of the mean 21-cm spectrum to parameter variations.


\bibliography{apssamp}
\bibliographystyle{abbrvnat-maxbibnames4}

\end{document}